%Paper: gr-qc/9412002
%From: bento@surya22.cern.ch (Maria Bento)
%Date: Thu, 1 Dec 1994 17:39:05 --100
%Date (revised): Thu, 1 Dec 1994 17:59:02 --100

%%%%%%%%%%%%%%%%%%%%%%%%%%%%%%%%%%%%%%%%%%%%%%%%%%%%%%%%%%%%%%%%%%%%%%%%%%%
%
% This file uses panda.tex (appended at the end)
%
%%%%%%%%%%%%%%%%%%%%%%%%%%%%%%%%%%%%%%%%%%%%%%%%%%%%%%%%%%%%%%%%%%%%%%%%%%%
\input panda
%\draftmode{string QC}
\loadamsmath
\pageno=0
\baselineskip=18pt
\nopagenumbers{
\line{\hfill CERN-TH.7488/94}
\line{\hfill DFTT 43/94}
\ifdoublepage \bjump\bjump\bjump\bjump\else\vfill\fi
\centerline{\bf Scale Factor Duality: A Quantum Cosmological Approach }
\bjump
\centerline{M. C. Bento\footnote{$^*$}{On leave of absence from Departamento de
F\' \i sica, Instituto
 Superior T\'ecnico, Av. Rovisco Pais, 1096 Lisboa Codex, Portugal}}
\centerline{\it CERN, Theory Division}
\centerline{\it CH-1211  Geneva 23}
\centerline{\it Switzerland}
\vskip 0.2cm
\centerline{ and}
\vskip 0.2cm
\centerline{O. Bertolami$^*$\footnote{$^\dagger$}{Also at Theory Division,
CERN.}}
\centerline{INFN-\it Sezione Torino}
\centerline{\it Via Pietro Giuria 1}
\centerline{\it I-10125  Torino, Italy}
\bjump
\ifdoublepage
\vfill
\noindent
\line{CERN-TH.7488/94\hfill}
\line{DFTT 43/94\hfill}
\line{October 1994\hfill}
\eject\null\vfill\fi
\centerline{\bf Abstract}\sjump
We consider the minisuperspace model arising from the lowest order
string effective action containing the graviton and the dilaton and
study solutions of the resulting Wheeler-DeWitt equation. The scale
factor duality symmetry is discussed in the context of our quantum
cosmological model.
\sjump\vfill
\ifdoublepage\else
\noindent
\line{CERN-TH.7488 /94\hfill}
\line{DFTT 43/94\hfill}
\line{November 1994\hfill}\fi
\eject}
\yespagenumbers\pageno=1

\def\ni{\noindent}
\def\laq{\raise 0.4ex\hbox{$<$}\kern -0.8em\lower 0.62 ex\hbox{$\sim$}}
\def\gaq{\raise 0.4ex\hbox{$>$}\kern -0.7em\lower 0.62 ex\hbox{$\sim$}}

%%%%%%%%%%%%%%%%%%%%%%%%%%%%%%%%%%%%%%%%%%%%%%%%%%%%%%%%%%%%%%%%%%%%%%%%%%%%%%
{\bf 1. Introduction}
\medskip
Duality transformations relate different, but actually equivalent,
conformal string backgrounds and, in particular, string theories
[\Ref{Schwarz}]. An example of this duality is the $O(d, d)$ transformation
connecting all toroidal compactifications in d-dimensions [\Ref{Narain}]. In
this latter case, it is shown that the duality transformation holds to
all orders in the string-loop expansion parameter by performing a suitable
change in the dilaton field when transforming metric and torsion
fields [\Ref{Alvarez}].
An important subset of this duality symmetries
is the so-called scale-factor or abelian duality  of string models
embedded in flat homogeneous and isotropic spacetimes
[\Ref{Veneziano}].
Scale-factor duality symmetry
is present in  the lowest order string effective action and means that the
transformation of the scale factor of a homogeneous and isotropic
target space metric, $a(t) \to a^{-1}(t)$, leaves the model invariant
provided that, in d spatial dimensions, the string coupling, i.e.
the dilaton, is properly transformed as well

$$\Phi(t)\rightarrow\varphi (t)=\Phi(t) - {d\over 2} \ln
a(t).
\nfr{a}

 Other transformations  were also  proposed to implement
these dualities  for backgrounds with non-abelian isometry
groups which are, in principle, compatible with homogeneous Bianchi
cosmological backgrounds [\Ref{Quevedo}] (see however [\Ref{Ricci},
\Ref{Gaume}]).

Scale-factor duality is an important guidance to introduce genuine stringy
features into an already known cosmological framework based on General
Relativity. However, although in the context of the resulting models one is
allowed to address important issues such as the problem of
singularities, inflation, generation of primordial energy density fluctuations
and show that a radiation dominated era naturally emerges from a
string cosmological scenario
(see e.g. Refs. [\Ref{Tseytlin}] and [\Ref{Gasperini}]),
these considerations and conclusions remain essentially
classical and are, therefore, presumably not capable of
capturing the deep quantum
gravity features one expects to extract from string theories.
Since a complete quantum string theory is not yet available,
a possible way to implement the classical stringy cosmological
scenarios discussed so far is to consider their quantum cosmological
extension by performing the canonical quantization of the
corresponding string model and solving the resulting Wheeler-DeWitt (WDW)
equation in the minisuperspace approximation.

In this work, we shall consider the
canonical quantization of the lowest order string effective action
using the standard ADM formalism in a $R\times S^3$ topology. This
choice allows us to remain within the standard formalism of quantum
cosmology and use its interpretative framework
[\Ref{Hartle}], although  scale factor
duality will be lost as an exact symmetry of the
resulting minisuperspace model.
\medskip
\ni
{\bf 2. The model and Wheeler-DeWitt equation}
\medskip
 We consider the following lowest order string effective action

$$S={1\over 2} \int_M d^D x \sqrt{-g} e^{-2\Phi} [R + 4 (\partial \Phi)^2 - 8
V(\Phi)],
\nfr{b}
where we have allowed for a potential for the dilaton field and we
have set the Kalb-Ramond tensor to vanish. We shall consider a
(D=4)--dimensional  homogeneous and isotropic spacetime described
by a closed Friedmann-Robertson-Walker metric.

In the canonical quantization formalism, the dynamical
piece of the metric  is the induced three-dimensional metric
$h_{ij}(i,j=1,2,3)$ on the boundary of the manifold $M$ over which
integration in \b\ is performed and, furthermore, to \b\ one has to add
the following boundary action:

$$S_B=-\int_{\partial M} d^3x \sqrt{h} e^{-2\Phi} K,
\nfr{c}
where $h$ is the determinant of the induced metric and K is the trace of
the second fundamental form on $\partial M$.

One obtains, after some computation, the following minisuperspace
Lagrangian density

$$ {\cal L}=N a^3 e^{-2\Phi} \left[ {3\over N^2} \left({\dot a \over
a}\right)^2 + {3\over a^2} -{6\over N^2} {\dot a\over a}\dot \Phi + 2
{{\dot \Phi}^2\over N^2} - 4 V(\Phi) \right],
\nfr{d}
where $N(t)$ is the lapse function.

Introducing the variable ${\displaystyle \tt z}(t)=\ln a(t)$ and the
transformation \a, one finds, in the N=1 gauge, that

$${\cal L}=e^{-2\varphi} \left[ -{3\over 2} {\dot {\displaystyle \tt
    z} }^2 + 3 e^{-2{\displaystyle \tt z}} + 2
{\dot \varphi}^2 -4 V(\varphi, {\displaystyle \tt z})\right].
\nfr{e}
which, except for the second term, exhibits the scale factor duality
symmetry under the transformation
${\displaystyle \tt z}\rightarrow -{\displaystyle \tt
z},\varphi\rightarrow\varphi$,
provided $V(\varphi,{\displaystyle \tt z})=V(\varphi,-{\displaystyle \tt z})$.

Aiming to obtain the WDW equation describing the quantum
cosmological features of  model \b, we compute the canonical
conjugate momenta to ${\displaystyle \tt z}$ and $\varphi$:

$$\eqalignno{\Pi_{\displaystyle \tt z} &={\partial {\cal L}\over
    \partial \dot {\displaystyle \tt z}}=-3 \dot {\displaystyle \tt z}
                       e^{-2\varphi}, & \nameali{ea}\cr
             \Pi_\varphi &={\partial {\cal L}\over \partial \dot \varphi}=4
\dot \varphi e^{-2\varphi}. & \nameali{eb}\cr}$$

The Hamiltonian density is then given by

$$ \eqalign{{\cal H} &=\Pi_{\displaystyle \tt z} \dot {\displaystyle
    \tt z} + \Pi_\varphi \dot \varphi - {\cal L}\cr
              &=e^{2\varphi} \left\{ -{1\over 6 }\Pi_{\displaystyle
                \tt z}^2 + {1\over 8}
\Pi_\varphi^2 + e^{-4\varphi} \left[ 4 V(\varphi,{\displaystyle \tt
  z}) - 3 e^{-2{\displaystyle \tt z}}\right]\right\},}
\nfr{f}
and  vanishes, on account of  invariance under time
reparametrization. The WDW equation is obtained
transforming this classical constraint, i.e.  ${\cal H}=0$,
into the vanishing of the Hamiltonian operator acting on the wavefunction
$\Psi(\varphi,{\displaystyle \tt z})$. For the latter step,
one has to promote the canonical conjugate momenta into operators:

$$\eqalign{\Pi_{\displaystyle \tt z} &\rightarrow -i {\partial\over
    \partial {\displaystyle \tt z}},\cr
           \Pi_\varphi &\rightarrow -i{\partial\over\partial\varphi}.}
\nfr{g}

Hence, one finds, after a proper operator ordering choice and trivial rescaling
of variables

$$\left[ {\partial^2\over\partial z^2}- {\partial^2\over\partial
\phi^2}+ U(\phi,z)\right]\Psi(\phi,z)=0,
\nfr{h}
where

$$U(\phi,z)={1\over 2}e^{-\phi} \left[ 4 V(\phi,z)- 3e^{-{1\over
\sqrt{3}}z}\right],
\nfr{i}
\noindent
and $\phi=4\varphi$, $z=2\sqrt{3}{\displaystyle \tt z}$.
%

%%%%%%%%%%%%%%%%%%%%%%%%%%%%%%%%%%%%%%%%%%%%%%%%%%%%%%%%%%%%%%%%%%%%%%%%%%%%%%
\medskip
\noindent
{\bf 3. Solutions of the Wheeler-DeWitt equation}
\medskip
We shall first study the case where the dilaton has no potential
(a scenario favoured e.g. in ref. [\Ref{Polyakov}])
and then proceed to analyse the effect of the introduction of a potential;
we shall consider the following simple cases:

$$\eqalignno{ V(\Phi)& =\Lambda, &\nameali{aa}  \cr
                    &  ={m^2\over 2} (\Phi-\Phi_o)^2, &\nameali{ab} \cr
                    &  =V_o e^{-\alpha \Phi}.&\nameali{ac}\cr}$$

The case of a cosmological constant, \aa, is the one which is compatible
with scale factor duality in a flat 3-dimensional spacetime [\Ref{Ritis}]. We
shall see, however, that the wave function  is fairly insensitive
to the 3-dimensional curvature and, using the interpretational rules
of quantum cosmology, that the most likely initial configuration
for the universe is
approximately compatible with scale factor duality. The second choice for the
dilaton potential  leads, in
the classical case, to chaotic inflationary solutions provided
$\Phi_i \gaq 4 M_P$, where $\Phi_i$ is the initial dilaton
configuration and $M_P$ is the Planck mass [\Ref{BBS},\Ref{BB}]; our quantum
analysis indicates that these configurations are actually
favoured. Finally, choice \ac\ is of  interest in connection
with  scenarios of supersymmetry breaking due to gaugino condensation
in the hidden sector of the theory [\Ref{Carlos}].

For the case where the dilaton has no potential,  the
corresponding WDW equation is  given by

$$ \left[ {\partial^2\over\partial z^2}- {\partial^2\over\partial
\phi^2}- {3\over 2}\ e^{-\phi-{1\over \sqrt{3}}z}\right]\Psi(\phi,z)=0.
\nfr{j}

Introducing the variables $x=\exp[-{1\over 2}(\phi + {1\over\sqrt{3}}z)]$
and $y=\phi+z$, this equation becomes

$$
\left[ x {\partial^2\over \partial x^2} - {1\over 2}  {\partial \over \partial
    x} - {A\over 2}{\partial^2 \over \partial x \partial y} + 9 x\right]
\Psi(x,y)=0,
\nfr{ja}
where $A=4(3-\sqrt{3})$. This equation  is separable, i.e. writing the
wave function as $\Psi(x,y)=F(x) G(y)$, one obtains

$$\eqalignno{{d^2 F\over d x^2} -  {1\over 2}(A\mu+1) {1\over x} {dF\over  d x
}+ 9 F & =0,
  & \nameali{jb}\cr
             {d G\over d y}-\mu G &=0, & \nameali{jc}\cr}$$
\noindent
where  $\mu$ is a separation constant. The solution of \j\ is then given by
[\Ref{AS}]:

$$\Psi(x,y)=c e^{\mu y} x^\alpha {\cal C}_{\alpha}(3  x),
\nfr{l}
where c is an integration constant,
$\alpha={1\over 4} [3 + A \mu]$ and ${\cal C }_{\alpha}$ is a generic
Bessel function of order $\alpha$. To ensure that $\lim_{a\to 0}\Psi =0$,
i.e. a singularity free cosmological scenario, we require that
$\mu > \mu_1={1\over 12(\sqrt{3} -1)}$ and, for $ \mu> 3\mu_1$,
the wave function is an increasing function in $\phi$.

 We shall now consider the situation where  the dilaton
potential is non-vanishing. Since the WDW equation
becomes more involved in this case and, within the spirit of
the quantum cosmology framework, we shall not look for  exact
solutions but rather  for the striking features of the wave function.
 We start with the simple
case of a constant dilaton potential, eq. \aa. In this case, the
minisuperspace potential is given by $U(\phi,z)={1\over 2}e^{-\phi}
\left[4\Lambda-3 e^{-{1\over\sqrt{3}}z}\right]$, which we shall analyse by
patching up solutions for $\phi>0$ (the wave function clearly vanishes
in the $\phi<0$ region due to the steepness of
the potential) in  the following three regions: $z>0$, $z<0$ and around
the origin. In fact, it is quite straightforward to show that
$\Psi$ essentially vanishes in the $z<0$ region.
For $z>0$, assuming that $4\Lambda\gg 3 e^{-{1\over\sqrt{3}}z}$ and that the
second term in $U(\phi,z)$ can therefore be neglected, the
resulting WDW equation becomes separable, i.e. the wave function can be
written as $\Psi(\phi,z)=F(z) G(\phi)$. It then yields the following
ordinary differential equations

$$\eqalignno{{d^2 F(z)\over dz^2 }-\mu F(z)&=0,  & \nameali{ma}\cr
{d^2 G(\phi)\over d \phi^2} - \left( 2 \Lambda e^{-\phi}
+ \mu \right) G(\phi) & =0, & \nameali{mb}\cr}$$

\noindent
where $\mu$ is the separation constant. The solution is given by [\Ref{AS}]

$$\eqalignno{\Psi(\phi,z)& =c e^{\pm \sqrt\mu z}
 {\cal Z}_{i\sqrt\mu}\left(\sqrt{2\Lambda} e^{-\phi/ 2}\right),
    \qquad        \hbox{ for }\ \mu >0,              &\nameali{oa}\cr
                         & =c \sin{\left(\sqrt|\mu|\ z\right)}
 {\cal  Z}_{\sqrt|\mu|}\left(\sqrt{2\Lambda} e^{-\phi/2}\right),     \qquad
   \hbox{ for }\ \mu <0,
    &\nameali{ob}\cr}$$

\noindent
for $\Lambda>0$, ${\cal Z}_\nu $ is a generic modified Bessel function of order
$\nu$. If $\Lambda<0$, ${\cal Z}_\nu$ should be replaced by ${\cal C}_\nu$
and $\Lambda$ by $|\Lambda|$ in the above equations.
Usual quantum cosmology
interpretational formalism is suitable for the ground-state wave
function of the universe [\Ref{Hartle}] and we shall consider henceforth the
case $\mu=0$. It then follows that the wave function, in the original
variables,  is given by

$$\Psi(\varphi,{\displaystyle \tt z}) =c {\displaystyle \tt z}
K_0\left(\sqrt{2\Lambda} e^{-2\varphi}\right),
\nfr{oc}

\noindent
for $\Lambda >0$, where $K_0$ is the modified Bessel function of order zero;
if $\Lambda <0$, $K_0$ should be replaced by the Bessel function
$J_0$ and $\Lambda$ by $|\Lambda|$.
Notice that, in both cases, the wave function is positive and
has a regular behaviour at the origin. Moreover, \oc\ implies that

$$\Psi\  \sim 2 \varphi\ \ln a,
    \qquad     \hbox{ for }\ {\varphi \to +\infty};
\nfr{od}

\noindent
hence $\Psi$ increases for large $\varphi$ and $a$.

Let us now study the region where $\phi$ and $z$ are close to the
origin. Expanding the exponentials up to second order and separating
variables, i.e. setting $\Psi(\phi,z)=F(z)G(\phi)$, we obtain  the
following ordinary differential equations

$$\eqalignno{{d^2 F(z)\over dz^2 }+ \left({\surd{3}\over 2} z + \mu\right)
F(z)&=0,  & \nameali{pa}\cr
{d^2 G(\phi)\over d \phi^2} + [\beta
(\phi - 1) - \mu]  G(\phi) & =0, & \nameali{pb}\cr}$$
\ni
 where $\beta=2\Lambda - 3/2  $. The solutions of these equations can
be given again  in terms of Bessel functions, respectively

$$\eqalignno{F(z)& =c {\hat z}^{1/2}{\cal
    C}_{1/3}\left(2\gamma^2\hat z^{3/2}
 \right)\qquad \hbox{ for }\  {\hat z}>0,     &\nameali{ra}\cr
                 & =c \left\vert \hat z\right\vert^{1/2}{\cal
    Z}_{1/3}\left(2\gamma^2  |\hat z|^{3/2}
 \right)\qquad \hbox{ for }\  {\hat z}<0,   &\nameali{rb}\cr}$$
where $\hat z={\surd{3}\over 2} z - \mu$,  $\gamma^2={2\surd 3\over 9}$ and

$$\eqalignno{G(\phi)& =c \hat \phi^{1/2}{\cal C}_{1/3}\left( {2\over
      3\beta} \hat \phi^{3/2}
 \right)\qquad \hbox{ for }\ \hat \phi>0,     &\nameali{raa}\cr
  &=c |\hat\phi|^{1/2}{\cal Z}_{1/3}\left( {2\over
      3\beta} |\hat\phi|^{3/2}
 \right)\qquad \hbox{ for }\ \hat\phi<0,   &\nameali{rba}\cr}$$
where $\hat \phi=\beta(\phi-1)-\mu$. For the vacuum state, $\mu=0$, we have

$$\eqalignno{\Psi(\phi, z)& = B (z\beta|\phi-1|)^{1/2} J_{1/3}
     (\gamma z^{3/2}) I_{1/3}\left( {2\over
      3} \beta^{1/2}|\phi-1|^{3/2}\right)~,
   \qquad                                  &\nameali{rc}\cr}$$

\noindent
for $\phi<1$, $\beta>0$ and $B$ a constant. If $\beta<0$,
$I_{1/3}$ should be interchanged by $J_{1/3}$ in the above equation
and $\beta$ replaced by $|\beta|$.
Hence, the ground-state wave function $\Psi(\phi, z)$
around the origin is essentially an increasing function of $\phi$ and $z$
[\Ref{AS}].

We find  that the salient features of the ground-state wave function are
that it is an increasing function of $\phi$ and $z$
and vanishing in the $\phi$ and
$z$ negative region. Hence, we conclude that
the favoured initial conditions are the ones for which $\phi$ and $z$ are
large.

We turn our attention to the case where the dilaton potential is
given by \ab. As discussed in Refs. [\Ref{BBS},\Ref{BB}], conditions for
successful chaotic inflation require $10^{-8}M_P<m<10^{-6}M_P$,
$\Phi_o\sim M_P$ and $ \Phi_i\gaq 4 M_P$.  For this choice of parameters,
the potential $U(\phi,z)$ is controlled by the overall exponential
factor, which implies that the previous results for the case of a
constant dilaton potential remain essentially unaltered. This means
that, as before, large values of $\phi$ and $z$ are the favoured  field
configurations.

Let us now turn to the case where the dilaton potential is given by
\ac. The value of $\alpha$ depends of the gauge group of the
hidden sector and for the known models $\alpha\gaq 24\pi^2/10$ [\Ref{Carlos}].
For these values of $\alpha$, the contribution of the dilaton potential
to $U(\phi,z)$ is  negligible and the ground-state wave
function is essentialy the one for the case where there is no  potential and
given by \l. Thus the conclusions drawn for that case are equally
valid here.
\medskip
\ni
{\bf 4. Conclusions}
\medskip
 Let us summarize our results and comment on some of its implications.
We have seen
that for any of the potentials we have analysed, including the case of
a vanishing potential,  the ground-state wave-function vanishes
for $z$ negative, which is consistent with the expectation that
$\Psi(a=0,\Phi)=0$ in order to solve the singularity problem. Another
generic feature of our results is that for all cases we have analysed
the ground-state
wave function of the universe is an increasing function of $\phi$. The
is true for the variable $z$ in the $z>0$ region.
 We can then conclude that the striking features of the wave function of the
universe are fairly independent on the dilaton potential, being also
insensitive to the presence of the spatial curvature.
Moreover, these conclusions imply that the favoured initial
field configurations are the
ones for which $\phi$ and $z$ are large (in Planck units) and that
the scale factor duality, although not a symmetry
of the classical closed FRW metric-dilaton system, actually holds as
an approximate symmetry. Thus, our results are compatible with
 conclusions drawn from
classical cosmological scenarios in which string features are
introduced through the scale factor duality  [\Ref{Tseytlin}]
and, in particular, with a pre-big-bang era [\Ref{Gasperini}]; furthermore,
our approach shows that quantization naturally solves the singularity problem.

 This approximate scale factor duality symmetry displayed by
the system we have analysed
implies that the ground-state wave
function of the universe will consist effectively of a superposition
of $\Psi(\Phi,a)$ and $\Psi(\Phi,a^{-1})$. It is interesting to speculate
on the possiblity that the transition from this superposed state to
the classical state, where  scale factor duality is lost, can be
achieved through the process of decoherence, with the inhomogeneous
modes of a massive dilaton field playing the role of the environment. For
this latter purpose, one could also consider massless dilaton and Yang-Mills
fields. For a fixed value  of the dilaton field, the ground-state wave
function for the Yang-Mills fields is essentially the one describing
an anharmonic oscillator with a quartic potential [\Ref{Bertold}].
Inhomogeneous modes of dilaton
and Yang-Mills fields could together drive decoherence  due to their
coupling. Decoherence could be also driven by massive vector fields [\Ref{BM}].

 Interestingly, one finds that, as in the case where one
considers only the dilaton and this field is endowed with a potential,
there exist chaotic inflationary solutions if $\Phi_i\gaq 4 M_P$, a
feature which holds even when including Yang-Mills fields [\Ref{BB}]. One
expects likewise that large initial values for the dilaton will also
be favoured in the quantum cosmological analysis of the
metric-Yang-Mills-dilaton system.
\bigskip
\ni
{\bf Acknowlegdements}
\indent

It is a pleasure to thank Dr. M. Cadoni and Profs. V. de Alfaro, M.
Gasperini and G. Veneziano for discussions.
\references
\beginref

\Rref{Schwarz} {J.H. Schwarz, in
``Elementary Particles and the Universe. Essays in honour of
M. Gell-Mann'', J. Schwarz ed. (Cambridge, 1991).}

\Rref{Alvarez} {E. Alvarez and M. Osorio, Phys. Rev. D40 (1989) 1150.}

\Rref{Veneziano}{G. Veneziano, Phys. Lett. B265 (1991) 287;
 A.A. Tseytlin, Mod. Phys. Lett. A6 (1991) 1721.}

\Rref{BB} {M.C. Bento and O. Bertolami, Phys. Lett. B336 (1994) 6.}

\Rref{BBS} {M.C. Bento, O. Bertolami and P.M. S\'a,
Phys. Lett. B262 (1991) 11; Mod. Phys. Lett. A7 (1992) 911.}

\Rref{Ritis} {S. Capozzielo and R. Ritis, Int. J. Mod. Phys. D2 (1993) 373.}

\Rref{Gasperini}{M. Gasperini and G. Veneziano,
Astropart. Phys. 1 (1993) 317.}

\Rref{Carlos}{B. de Carlos, J.A. Casas and C. Mu\~noz, Nucl. Phys. B339
   (1993) 623.}

\Rref{Hartle} {J.B. Hartle and S.W. Hawking,
Phys. Rev. D28 (1983) 2960;
J.B. Hartle, in Astrophysics, Cargese 1986, B. Carter and J.B. Hartle
eds. (Plenum, New York, 1986).}

\Rref{Polyakov} {T. Damour and A.M. Polyakov,  Nucl. Phys. B423 (1994) 532. }

\Rref{Bertold} {O. Bertolami and J.M. Mour\~ao,
Class. Quantum Gravity 8 (1991) 1271.}

\Rref{BM} {O. Bertolami and P.V. Moniz, ``Decoherence of
  Friedmann-Robertson-Walker Geometries in the Presence of Massive
  Vector Fields'', Preprint CERN-TH 7241/94, DAMTP R-94/22.}

\Rref{Narain} {K.S. Narain, Phys. Lett. B169 (1986) 41;
 K.S. Narain, M.H. Sarmadi and E. Witten,
              Nucl. Phys. B279(1987)369.}

\Rref{Quevedo} {X.C. de la Ossa and F. Quevedo,
Nucl. Phys. B403 (1993) 377.}

\Rref{Ricci}{M. Gasperini, R. Ricci and G. Veneziano, Phys. Lett. B319
  (1993) 438.}

\Rref{Gaume} {E. Alvarez, L. Alvarez-Gaum\'e, J.L.F. Barb\'on and Y.  Lozano,
Nucl. Phys. B415 (1994) 71.}

\Rref{Tseytlin} {A.A. Tseytlin and C. Vafa, Nucl. Phys. B372 (1992) 443.}

\Rref{AS} {M. Abramowitz and I.A. Stegun, ``Handbook of Mathematical
Functions'' (Dover, New York, 1970). }
\endref
\ciao

%%%%%%%%%%%%%%%%%%%%%%%%%%%%%%%%%%%%%%%%%%%%%%%%%%%%%%%%%%%%%%%%%%%%%
%
%
%                                      Andrea PASQUINUCCI, 1988
%              PANDA.TEX               S.I.S.S.A., Trieste, Italy
%                                      (Revised 1991, Princeton, USA)
%
%--------------------------------------------------------------------
%
%    These are TEX macros. They work with PLAIN TEX (the basis
%    version of TEX). The only problem can be with the double-page
%    format since it depends on the type of software and laserwriter
%    you use to print, so I cannot guarantee that the double-page
%    format will work properly. Double-page MUST be printed in
%    LANDSCAPE orientation. (You shouldn't have troubles with fonts;
%    if you do, please let me know.)
%
%--------------------------------------------------------------------
%
%                     INTERACTIVE SECTION
%
%--------------------------------------------------------------------
%
\def\standardrisposta{s }\def\reducedrisposta{r }
\def\mplarisposta{mpla }\def\zerorisposta{z }
\def\doublerisposta{d }\def\cartarisposta{e }\def\amsrisposta{y }
\newcount\ingrandimento \newcount\sinnota \newcount\dimnota
\newcount\unoduecol \newdimen\collhsize \newdimen\tothsize
\newdimen\fullhsize \newcount\controllorisposta \sinnota=1
\newskip\infralinea  \global\controllorisposta=0
\immediate\write16 { ********  Welcome to PANDA macros (Plain TeX,
AP, 1991) ******** }
\immediate\write16 { You'll have to answer a few questions in
lowercase.}
\message{>  Do you want it in double-page (d), reduced (r)
or standard format (s) ? }\read-1 to\risposta
\message{>  Do you want it in USA A4 (u) or EUROPEAN A4
(e) paper size ? }\read-1 to\srisposta
%\message{>  Do you have AMSFonts 2.0 (math) fonts (y/n) ? }
%\read-1 to\arisposta
%
%--------------------------------------------------------------------
%
%             END INTERACTIVE SECTION - PAGE FORMATTING
%
%--------------------------------------------------------------------
%       The following parameters define defaults to the interactive
%       session.  At the moment I have set EUROPEAN and MATH FONTS
%\def\risposta{d } \def\srisposta{u } \def\arisposta{n }
\def\srisposta{e }
\def\arisposta{y }
\ifx\risposta\standardrisposta \ingrandimento=1200
\message {>> This will come out UNREDUCED << }
\dimnota=2 \unoduecol=1 \global\controllorisposta=1 \fi
\ifx\risposta\reducedrisposta \ingrandimento=1095 \dimnota=1
\unoduecol=1  \global\controllorisposta=1
\message {>> This will come out REDUCED << } \fi
\ifx\risposta\doublerisposta \ingrandimento=1000 \dimnota=2
\unoduecol=2   \message {>> You must print this in
LANDSCAPE orientation << } \global\controllorisposta=1 \fi
\ifx\risposta\mplarisposta \ingrandimento=1000 \dimnota=1
\message {>> Mod. Phys. Lett. A format << }
\unoduecol=1 \global\controllorisposta=1 \fi
\ifx\risposta\zerorisposta \ingrandimento=1000 \dimnota=2
\message {>> Zero Magnification format << }
\unoduecol=1 \global\controllorisposta=1 \fi
\ifnum\controllorisposta=0  \ingrandimento=1200
\message {>>> ERROR IN INPUT, I ASSUME STANDARD
UNREDUCED FORMAT <<< }  \dimnota=2 \unoduecol=1 \fi
\magnification=\ingrandimento
%
%--------------------------------------------------------------------
%
%                        PARAMETERS SETTING
%
%  You can modify these parameters at your will (and resposability)
%--------------------------------------------------------------------
%
\newdimen\eucolumnsize \newdimen\eudoublehsize \newdimen\eudoublevsize
\newdimen\uscolumnsize \newdimen\usdoublehsize \newdimen\usdoublevsize
\newdimen\eusinglehsize \newdimen\eusinglevsize \newdimen\ussinglehsize
\newskip\standardbaselineskip \newdimen\ussinglevsize
\newskip\reducedbaselineskip \newskip\doublebaselineskip
\eucolumnsize=12.0truecm    % column h-size for european doublepage
                            % (12.0treucm default)
\eudoublehsize=25.5truecm   % sheet h-size for european duoblepage
                            % (25.5treucm default)
\eudoublevsize=6.7truein    % sheet v-size for european doublepage
                            % (6.5treuin default  or 17truecm?)
\uscolumnsize=4.4truein     % column h-size for american doublepage
                            % (4.4treuin default)
\usdoublehsize=9.4truein    % sheet h-size for american duoblepage
                            % (9.4treuin default)
\usdoublevsize=6.8truein    % sheet v-size for american doublepage
                            % (6.8treuin default)
\eusinglehsize=6.5truein    % sheet h-size for european singlepage
                            % (6.5truein default)
\eusinglevsize=24truecm     % sheet v-size for european singlepage
                            % (24truecm default)
\ussinglehsize=6.5truein    % sheet h-size for american singlepage
                            % (6.5truein default)
\ussinglevsize=8.9truein    % sheet v-size for american singlepage
                            % (8.9truein default)
\standardbaselineskip=16pt plus.2pt  % baselineskip for standard
                                     % format (16pt default)
\reducedbaselineskip=14pt plus.2pt   % baselineskip for reduced
                                     % format (14pt default)
\doublebaselineskip=12pt plus.2pt    % baselineskip for doublepage
                                     % format (12pt default)
%
%  \Portoffset and \Landoffset define the horizontal and vertical
%  offsets respectively for portrait and landscape modes. Example:
%  \def\Portoffset{\voffset=.4truein\hoffset=.125truein}
%
\def\Portoffset{}
\def\Landoffset{\voffset=-.2truein}
\ifx\risposta\mplarisposta \def\Portoffset{\hoffset=1.8truecm} \fi
%
%  \Landspec defines the \special command that sets the printer
%  to landscape mode without need to specify it directly in the
%  TeX to postscript translator (the command is site dependent).
%  Example: \def\Landspec{\special{ps: landscape}}
%
\def\Landspec{}
\tolerance=10000
\parskip=0pt plus2pt  \leftskip=0pt \rightskip=0pt
%
%   Do not modify anything of what follows
%                       (unless you know what you are doing!)
%----------------------------------------------------------------------
%
\ifx\risposta\standardrisposta \infralinea=\standardbaselineskip \fi
\ifx\risposta\reducedrisposta  \infralinea=\reducedbaselineskip \fi
\ifx\risposta\doublerisposta   \infralinea=\doublebaselineskip \fi
\ifx\risposta\mplarisposta     \infralinea=13pt \fi
\ifx\risposta\zerorisposta     \infralinea=12pt plus.2pt\fi
\ifnum\controllorisposta=0    \infralinea=\standardbaselineskip \fi
\ifx\risposta\doublerisposta   \Landoffset \else \Portoffset \fi
\ifx\risposta\doublerisposta \ifx\srisposta\cartarisposta
\tothsize=\eudoublehsize \collhsize=\eucolumnsize
\vsize=\eudoublevsize  \else  \tothsize=\usdoublehsize
\collhsize=\uscolumnsize \vsize=\usdoublevsize \fi \else
\ifx\srisposta\cartarisposta \tothsize=\eusinglehsize
\vsize=\eusinglevsize \else  \tothsize=\ussinglehsize
\vsize=\ussinglevsize \fi \collhsize=4.4truein \fi
\ifx\risposta\mplarisposta \tothsize=5.0truein
\vsize=7.8truein \collhsize=4.4truein \fi
%
%--------------------------------------------------------------------
%
%                            FONTS
%
%--------------------------------------------------------------------
%
\newcount\contaeuler \newcount\contacyrill \newcount\contaams
\font\ninerm=cmr9  \font\eightrm=cmr8  \font\sixrm=cmr6
\font\ninei=cmmi9  \font\eighti=cmmi8  \font\sixi=cmmi6
\font\ninesy=cmsy9  \font\eightsy=cmsy8  \font\sixsy=cmsy6
\font\ninebf=cmbx9  \font\eightbf=cmbx8  \font\sixbf=cmbx6
\font\ninett=cmtt9  \font\eighttt=cmtt8  \font\nineit=cmti9
\font\eightit=cmti8 \font\ninesl=cmsl9  \font\eightsl=cmsl8
\skewchar\ninei='177 \skewchar\eighti='177 \skewchar\sixi='177
\skewchar\ninesy='60 \skewchar\eightsy='60 \skewchar\sixsy='60
\hyphenchar\ninett=-1 \hyphenchar\eighttt=-1 \hyphenchar\tentt=-1
%
                 % math italic bold \bfmath
\font\tencmmib=cmmib10  \newfam\cmmibfam  \skewchar\tencmmib='177
                  % math bold (cal) symbols
\font\tencmbsy=cmbsy10  \newfam\cmbsyfam  \skewchar\tencmbsy='60
                 % small caps (uppercase)
\font\tencmcsc=cmcsc10  \newfam\cmcscfam
\ifnum\ingrandimento=1095

\else

\fi

\def\ttaarr{\bf}                % chapter titles' font
\def\ppaarr{\sl}                % section titles' font

%
     % inch-high caps (enormous)
%
%   AMS fonts (this works only if you have at least the 2.0
%              version of AMSFonts, otherwise say no)
%
\newfam\eufmfam \newfam\msamfam \newfam\msbmfam \newfam\eufbfam
\def\Loadeulerfonts{\global\contaeuler=1 \ifx\arisposta\amsrisposta
\font\teneufm=eufm10              %  \eufm   Gothic (or Euler)
\font\eighteufm=eufm8 \font\nineeufm=eufm9 \font\sixeufm=eufm6
\font\seveneufm=eufm7  \font\fiveeufm=eufm5
\font\teneufb=eufb10              %  \eufb   Bold Gothic (or Euler)
\font\eighteufb=eufb8 \font\nineeufb=eufb9 \font\sixeufb=eufb6
\font\seveneufb=eufb7  \font\fiveeufb=eufb5
\font\teneurm=eurm10              %  \eurm   Roman Gothic (or Euler)
\font\eighteurm=eurm8 \font\nineeurm=eurm9
\font\teneurb=eurb10              %  \eurb   Roman Bold Gothic
\font\eighteurb=eurb8 \font\nineeurb=eurb9
\font\teneusm=eusm10              %  \eusm   Slanted Capital Gothic
\font\eighteusm=eusm8 \font\nineeusm=eusm9
\font\teneusb=eusb10              %\eusb Slanted Capital Bold Gothic
\font\eighteusb=eusb8 \font\nineeusb=eusb9
\else \def\eufm{\tt} \def\eufb{\tt} \def\eurm{\tt} \def\eurb{\tt}
\def\eusm{\tt} \def\eusb{\tt}    \fi}

\def\loadamsmath{\global\contaams=1 \ifx\arisposta\amsrisposta
\font\tenmsam=msam10 \font\ninemsam=msam9 \font\eightmsam=msam8
\font\sevenmsam=msam7 \font\sixmsam=msam6 \font\fivemsam=msam5
\font\tenmsbm=msbm10 \font\ninemsbm=msbm9 \font\eightmsbm=msbm8
\font\sevenmsbm=msbm7 \font\sixmsbm=msbm6 \font\fivemsbm=msbm5
\else \def\msbm{\bf} \fi \def\Bbb{\msbm} \def\symbl{\msam} \tenpoint}
\def\loadcyrill{\global\contacyrill=1 \ifx\arisposta\amsrisposta
\font\tenwncyr=wncyr10 \font\ninewncyr=wncyr9 \font\eightwncyr=wncyr8
\font\tenwncyb=wncyr10 \font\ninewncyb=wncyr9 \font\eightwncyb=wncyr8
\font\tenwncyi=wncyr10 \font\ninewncyi=wncyr9 \font\eightwncyi=wncyr8
\else \def\cyrill{\sl} \def\cyrilb{\sl} \def\cyrili{\sl} \fi\tenpoint}
\ifx\arisposta\amsrisposta
\font\sevenex=cmex7               %  reduced math symbols
\font\eightex=cmex8  \font\nineex=cmex9
\font\ninecmmib=cmmib9   \font\eightcmmib=cmmib8
\font\sevencmmib=cmmib7 \font\sixcmmib=cmmib6
\font\fivecmmib=cmmib5   \skewchar\ninecmmib='177
\skewchar\eightcmmib='177  \skewchar\sevencmmib='177
\skewchar\sixcmmib='177   \skewchar\fivecmmib='177
\font\ninecmbsy=cmbsy9    \font\eightcmbsy=cmbsy8
\font\sevencmbsy=cmbsy7  \font\sixcmbsy=cmbsy6
\font\fivecmbsy=cmbsy5   \skewchar\ninecmbsy='60
\skewchar\eightcmbsy='60  \skewchar\sevencmbsy='60
\skewchar\sixcmbsy='60    \skewchar\fivecmbsy='60
\font\ninecmcsc=cmcsc9    \font\eightcmcsc=cmcsc8     \else
\def\cmmib{\fam\cmmibfam\tencmmib}\textfont\cmmibfam=\tencmmib
\scriptfont\cmmibfam=\tencmmib \scriptscriptfont\cmmibfam=\tencmmib
\def\cmbsy{\fam\cmbsyfam\tencmbsy} \textfont\cmbsyfam=\tencmbsy
\scriptfont\cmbsyfam=\tencmbsy \scriptscriptfont\cmbsyfam=\tencmbsy
\scriptfont\cmcscfam=\tencmcsc \scriptscriptfont\cmcscfam=\tencmcsc
\def\cmcsc{\fam\cmcscfam\tencmcsc} \textfont\cmcscfam=\tencmcsc \fi
\catcode`@=11
\newskip\ttglue
\gdef\tenpoint{\def\rm{\fam0\tenrm}
  \textfont0=\tenrm \scriptfont0=\sevenrm \scriptscriptfont0=\fiverm
  \textfont1=\teni \scriptfont1=\seveni \scriptscriptfont1=\fivei
  \textfont2=\tensy \scriptfont2=\sevensy \scriptscriptfont2=\fivesy
  \textfont3=\tenex \scriptfont3=\tenex \scriptscriptfont3=\tenex
  \def\mcal{\fam2 \tensy}  \def\mmit{\fam1 \teni}
  \textfont\itfam=\tenit \def\it{\fam\itfam\tenit}
  \textfont\slfam=\tensl \def\sl{\fam\slfam\tensl}
  \textfont\ttfam=\tentt \scriptfont\ttfam=\eighttt
  \scriptscriptfont\ttfam=\eighttt  \def\tt{\fam\ttfam\tentt}
  \textfont\bffam=\tenbf \scriptfont\bffam=\sevenbf
  \scriptscriptfont\bffam=\fivebf \def\bf{\fam\bffam\tenbf}
     \ifx\arisposta\amsrisposta    \ifnum\contaeuler=1
  \textfont\eufmfam=\teneufm \scriptfont\eufmfam=\seveneufm
  \scriptscriptfont\eufmfam=\fiveeufm \def\eufm{\fam\eufmfam\teneufm}
  \textfont\eufbfam=\teneufb \scriptfont\eufbfam=\seveneufb
  \scriptscriptfont\eufbfam=\fiveeufb \def\eufb{\fam\eufbfam\teneufb}
  \def\eurm{\teneurm} \def\eurb{\teneurb} \def\eusm{\teneusm}
  \def\eusb{\teneusb}    \fi    \ifnum\contaams=1
  \textfont\msamfam=\tenmsam \scriptfont\msamfam=\sevenmsam
  \scriptscriptfont\msamfam=\fivemsam \def\msam{\fam\msamfam\tenmsam}
  \textfont\msbmfam=\tenmsbm \scriptfont\msbmfam=\sevenmsbm
  \scriptscriptfont\msbmfam=\fivemsbm \def\msbm{\fam\msbmfam\tenmsbm}
     \fi      \ifnum\contacyrill=1     \def\cyrill{\tenwncyr}
  \def\cyrilb{\tenwncyb}  \def\cyrili{\tenwncyi}         \fi
  \textfont3=\tenex \scriptfont3=\sevenex \scriptscriptfont3=\sevenex
  \def\cmmib{\fam\cmmibfam\tencmmib} \scriptfont\cmmibfam=\sevencmmib
  \textfont\cmmibfam=\tencmmib  \scriptscriptfont\cmmibfam=\fivecmmib
  \def\cmbsy{\fam\cmbsyfam\tencmbsy} \scriptfont\cmbsyfam=\sevencmbsy
  \textfont\cmbsyfam=\tencmbsy  \scriptscriptfont\cmbsyfam=\fivecmbsy
  \def\cmcsc{\fam\cmcscfam\tencmcsc} \scriptfont\cmcscfam=\eightcmcsc
  \textfont\cmcscfam=\tencmcsc \scriptscriptfont\cmcscfam=\eightcmcsc
     \fi            \tt \ttglue=.5em plus.25em minus.15em
  \normalbaselineskip=12pt
  \setbox\strutbox=\hbox{\vrule height8.5pt depth3.5pt width0pt}
  \let\sc=\eightrm \let\big=\tenbig   \normalbaselines
  \baselineskip=\infralinea  \rm}
\gdef\ninepoint{\def\rm{\fam0\ninerm}
  \textfont0=\ninerm \scriptfont0=\sixrm \scriptscriptfont0=\fiverm
  \textfont1=\ninei \scriptfont1=\sixi \scriptscriptfont1=\fivei
  \textfont2=\ninesy \scriptfont2=\sixsy \scriptscriptfont2=\fivesy
  \textfont3=\tenex \scriptfont3=\tenex \scriptscriptfont3=\tenex
  \def\mcal{\fam2 \ninesy}  \def\mmit{\fam1 \ninei}
  \textfont\itfam=\nineit \def\it{\fam\itfam\nineit}
  \textfont\slfam=\ninesl \def\sl{\fam\slfam\ninesl}
  \textfont\ttfam=\ninett \scriptfont\ttfam=\eighttt
  \scriptscriptfont\ttfam=\eighttt \def\tt{\fam\ttfam\ninett}
  \textfont\bffam=\ninebf \scriptfont\bffam=\sixbf
  \scriptscriptfont\bffam=\fivebf \def\bf{\fam\bffam\ninebf}
     \ifx\arisposta\amsrisposta  \ifnum\contaeuler=1
  \textfont\eufmfam=\nineeufm \scriptfont\eufmfam=\sixeufm
  \scriptscriptfont\eufmfam=\fiveeufm \def\eufm{\fam\eufmfam\nineeufm}
  \textfont\eufbfam=\nineeufb \scriptfont\eufbfam=\sixeufb
  \scriptscriptfont\eufbfam=\fiveeufb \def\eufb{\fam\eufbfam\nineeufb}
  \def\eurm{\nineeurm} \def\eurb{\nineeurb} \def\eusm{\nineeusm}
  \def\eusb{\nineeusb}     \fi   \ifnum\contaams=1
  \textfont\msamfam=\ninemsam \scriptfont\msamfam=\sixmsam
  \scriptscriptfont\msamfam=\fivemsam \def\msam{\fam\msamfam\ninemsam}
  \textfont\msbmfam=\ninemsbm \scriptfont\msbmfam=\sixmsbm
  \scriptscriptfont\msbmfam=\fivemsbm \def\msbm{\fam\msbmfam\ninemsbm}
     \fi       \ifnum\contacyrill=1     \def\cyrill{\ninewncyr}
  \def\cyrilb{\ninewncyb}  \def\cyrili{\ninewncyi}         \fi
  \textfont3=\nineex \scriptfont3=\sevenex \scriptscriptfont3=\sevenex
  \def\cmmib{\fam\cmmibfam\ninecmmib}  \textfont\cmmibfam=\ninecmmib
  \scriptfont\cmmibfam=\sixcmmib \scriptscriptfont\cmmibfam=\fivecmmib
  \def\cmbsy{\fam\cmbsyfam\ninecmbsy}  \textfont\cmbsyfam=\ninecmbsy
  \scriptfont\cmbsyfam=\sixcmbsy \scriptscriptfont\cmbsyfam=\fivecmbsy
  \def\cmcsc{\fam\cmcscfam\ninecmcsc} \scriptfont\cmcscfam=\eightcmcsc
  \textfont\cmcscfam=\ninecmcsc \scriptscriptfont\cmcscfam=\eightcmcsc
     \fi            \tt \ttglue=.5em plus.25em minus.15em
  \normalbaselineskip=11pt
  \setbox\strutbox=\hbox{\vrule height8pt depth3pt width0pt}
  \let\sc=\sevenrm \let\big=\ninebig \normalbaselines\rm}
\gdef\eightpoint{\def\rm{\fam0\eightrm}
  \textfont0=\eightrm \scriptfont0=\sixrm \scriptscriptfont0=\fiverm
  \textfont1=\eighti \scriptfont1=\sixi \scriptscriptfont1=\fivei
  \textfont2=\eightsy \scriptfont2=\sixsy \scriptscriptfont2=\fivesy
  \textfont3=\tenex \scriptfont3=\tenex \scriptscriptfont3=\tenex
  \def\mcal{\fam2 \eightsy}  \def\mmit{\fam1 \eighti}
  \textfont\itfam=\eightit \def\it{\fam\itfam\eightit}
  \textfont\slfam=\eightsl \def\sl{\fam\slfam\eightsl}
  \textfont\ttfam=\eighttt \scriptfont\ttfam=\eighttt
  \scriptscriptfont\ttfam=\eighttt \def\tt{\fam\ttfam\eighttt}
  \textfont\bffam=\eightbf \scriptfont\bffam=\sixbf
  \scriptscriptfont\bffam=\fivebf \def\bf{\fam\bffam\eightbf}
     \ifx\arisposta\amsrisposta   \ifnum\contaeuler=1
  \textfont\eufmfam=\eighteufm \scriptfont\eufmfam=\sixeufm
  \scriptscriptfont\eufmfam=\fiveeufm \def\eufm{\fam\eufmfam\eighteufm}
  \textfont\eufbfam=\eighteufb \scriptfont\eufbfam=\sixeufb
  \scriptscriptfont\eufbfam=\fiveeufb \def\eufb{\fam\eufbfam\eighteufb}
  \def\eurm{\eighteurm} \def\eurb{\eighteurb} \def\eusm{\eighteusm}
  \def\eusb{\eighteusb}       \fi    \ifnum\contaams=1
  \textfont\msamfam=\eightmsam \scriptfont\msamfam=\sixmsam
  \scriptscriptfont\msamfam=\fivemsam \def\msam{\fam\msamfam\eightmsam}
  \textfont\msbmfam=\eightmsbm \scriptfont\msbmfam=\sixmsbm
  \scriptscriptfont\msbmfam=\fivemsbm \def\msbm{\fam\msbmfam\eightmsbm}
     \fi       \ifnum\contacyrill=1     \def\cyrill{\eightwncyr}
  \def\cyrilb{\eightwncyb}  \def\cyrili{\eightwncyi}         \fi
  \textfont3=\eightex \scriptfont3=\sevenex \scriptscriptfont3=\sevenex
  \def\cmmib{\fam\cmmibfam\eightcmmib}  \textfont\cmmibfam=\eightcmmib
  \scriptfont\cmmibfam=\sixcmmib \scriptscriptfont\cmmibfam=\fivecmmib
  \def\cmbsy{\fam\cmbsyfam\eightcmbsy}  \textfont\cmbsyfam=\eightcmbsy
  \scriptfont\cmbsyfam=\sixcmbsy \scriptscriptfont\cmbsyfam=\fivecmbsy
  \def\cmcsc{\fam\cmcscfam\eightcmcsc} \scriptfont\cmcscfam=\eightcmcsc
  \textfont\cmcscfam=\eightcmcsc \scriptscriptfont\cmcscfam=\eightcmcsc
     \fi             \tt \ttglue=.5em plus.25em minus.15em
  \normalbaselineskip=9pt
  \setbox\strutbox=\hbox{\vrule height7pt depth2pt width0pt}
  \let\sc=\sixrm \let\big=\eightbig \normalbaselines\rm }
\gdef\tenbig#1{{\hbox{$\left#1\vbox to8.5pt{}\right.\n@space$}}}
\gdef\ninebig#1{{\hbox{$\textfont0=\tenrm\textfont2=\tensy
   \left#1\vbox to7.25pt{}\right.\n@space$}}}
\gdef\eightbig#1{{\hbox{$\textfont0=\ninerm\textfont2=\ninesy
   \left#1\vbox to6.5pt{}\right.\n@space$}}}
 %for 10-pt math in 9-pt territory
\def\alternativefont#1#2{\ifx\arisposta\amsrisposta \relax \else
\xdef#1{#2} \fi}
\global\contaeuler=0 \global\contacyrill=0 \global\contaams=0
%
%--------------------------------------------------------------------
%
%                            MACROS
%
%--------------------------------------------------------------------
%
\newbox\fotlinebb \newbox\hedlinebb \newbox\leftcolumn
\gdef\makeheadline{\vbox to 0pt{\vskip-22.5pt
     \fullline{\vbox to8.5pt{}\the\headline}\vss}\nointerlineskip}
\gdef\makehedlinebb{\vbox to 0pt{\vskip-22.5pt
     \fullline{\vbox to8.5pt{}\copy\hedlinebb\hfil
     \line{\hfill\the\headline\hfill}}\vss} \nointerlineskip}
\gdef\makefootline{\baselineskip=24pt \fullline{\the\footline}}
\gdef\makefotlinebb{\baselineskip=24pt
    \fullline{\copy\fotlinebb\hfil\line{\hfill\the\footline\hfill}}}
\gdef\doubleformat{\shipout\vbox{\Landspec\makehedlinebb
     \fullline{\box\leftcolumn\hfil\columnbox}\makefotlinebb}
     \advancepageno}
\gdef\columnbox{\leftline{\pagebody}}
\gdef\line#1{\hbox to\hsize{\hskip\leftskip#1\hskip\rightskip}}
\gdef\fullline#1{\hbox to\fullhsize{\hskip\leftskip{#1}%
\hskip\rightskip}}
\gdef\footnote#1{\let\@sf=\empty
         \ifhmode\edef\#sf{\spacefactor=\the\spacefactor}\/\fi
         #1\@sf\vfootnote{#1}}
\gdef\vfootnote#1{\insert\footins\bgroup
         \ifnum\dimnota=1  \eightpoint\fi
         \ifnum\dimnota=2  \ninepoint\fi
         \ifnum\dimnota=0  \tenpoint\fi
         \interlinepenalty=\interfootnotelinepenalty
         \splittopskip=\ht\strutbox
         \splitmaxdepth=\dp\strutbox \floatingpenalty=20000
         \leftskip=\oldssposta \rightskip=\olddsposta
         \spaceskip=0pt \xspaceskip=0pt
         \ifnum\sinnota=0   \textindent{#1}\fi
         \ifnum\sinnota=1   \item{#1}\fi
         \footstrut\futurelet\next\fo@t}
\gdef\fo@t{\ifcat\bgroup\noexpand\next \let\next\f@@t
             \else\let\next\f@t\fi \next}
\gdef\f@@t{\bgroup\aftergroup\@foot\let\next}
\gdef\f@t#1{#1\@foot} \gdef\@foot{\strut\egroup}
\gdef\footstrut{\vbox to\splittopskip{}}
\skip\footins=\bigskipamount
\count\footins=1000  \dimen\footins=8in
\catcode`@=12
\tenpoint
\ifnum\unoduecol=1 \hsize=\tothsize   \fullhsize=\tothsize \fi
\ifnum\unoduecol=2 \hsize=\collhsize  \fullhsize=\tothsize \fi
\global\let\lrcol=L      \ifnum\unoduecol=1
\output{\plainoutput{\ifnum\tipbnota=2 \clearnmbnota\fi}} \fi
\ifnum\unoduecol=2 \output{\if L\lrcol
     \global\setbox\leftcolumn=\columnbox
     \global\setbox\fotlinebb=\line{\hfill\the\footline\hfill}
     \global\setbox\hedlinebb=\line{\hfill\the\headline\hfill}
     \advancepageno  \global\let\lrcol=R
     \else  \doubleformat \global\let\lrcol=L \fi
     \ifnum\outputpenalty>-20000 \else\dosupereject\fi
     \ifnum\tipbnota=2\clearnmbnota\fi }\fi
\def\ifdoublepage{\ifnum\unoduecol=2 }
\gdef\yespagenumbers{\footline={\hss\tenrm\folio\hss}}
\gdef\ciao{ \ifnum\fdefcontre=1 \endfdef\fi
     \par\vfill\supereject \ifnum\unoduecol=2
     \if R\lrcol  \headline={}\nopagenumbers\null\vfill\eject
     \fi\fi \end}

\newskip\olddsposta \newskip\oldssposta
\global\oldssposta=\leftskip \global\olddsposta=\rightskip

\def\filldots{\leaders\hbox to 1em{\hss.\hss}\hfill}
\def\inquadrb#1 {\vbox {\hrule  \hbox{\vrule \vbox {\vskip .2cm
    \hbox {\ #1\ } \vskip .2cm } \vrule  }  \hrule} }
 \def\newline{\hfil\break}
\def\jump{\vskip\baselineskip} \newskip\iinnffrr
\def\sjump{\iinnffrr=\baselineskip
          \divide\iinnffrr by 2 \vskip\iinnffrr}
\def\bjump{\vskip\baselineskip \vskip\baselineskip}
\newcount\nmbnota  \def\clearnmbnota{\global\nmbnota=0}
\newcount\tipbnota \def\letterfootnote{\global\tipbnota=1}

\def\note#1{\global\advance\nmbnota by 1 \ifnum\tipbnota=1
    \footnote{$^{\rm\nttlett}$}{#1} \else {\ifnum\tipbnota=2
    \footnote{$^{\nttsymb}$}{#1}
    \else\footnote{$^{\the\nmbnota}$}{#1}\fi}\fi}
\def\nttlett{\ifcase\nmbnota \or a\or b\or c\or d\or e\or f\or
g\or h\or i\or j\or k\or l\or m\or n\or o\or p\or q\or r\or
s\or t\or u\or v\or w\or y\or x\or z\fi}
\def\nttsymb{\ifcase\nmbnota \or\dag\or\sharp\or\ddag\or\star\or
\natural\or\flat\or\clubsuit\or\diamondsuit\or\heartsuit
\or\spadesuit\fi}   \clearnmbnota
\def\numberfootnote{\global\tipbnota=0} \numberfootnote
\def\setnote#1{\expandafter\xdef\csname#1\endcsname{
\ifnum\tipbnota=1 {\rm\nttlett} \else {\ifnum\tipbnota=2
{\nttsymb} \else \the\nmbnota\fi}\fi} }
\newcount\nbmfig  \def\clearnbmfig{\global\nbmfig=0}
\gdef\figure{\global\advance\nbmfig by 1
      {\rm fig. \the\nbmfig}}   \clearnbmfig
\def\setfig#1{\expandafter\xdef\csname#1\endcsname{fig. \the\nbmfig}}

\newcount\frmcount \def\clearfrmcount{\global\frmcount=0}
\def\numero{\global\advance\frmcount by 1   \ifnum\indappcount=0
  {\ifnum\cpcount <1 {\hbox{\rm (\the\frmcount )}}  \else
  {\hbox{\rm (\the\cpcount .\the\frmcount )}} \fi}  \else
  {\hbox{\rm (\applett .\the\frmcount )}} \fi}
\def\nameformula#1{\global\advance\frmcount by 1%
\ifnum\draftnum=0  {\ifnum\indappcount=0%
{\ifnum\cpcount<1\xdef\spzzttrra{(\the\frmcount )}%
\else\xdef\spzzttrra{(\the\cpcount .\the\frmcount )}\fi}%
\else\xdef\spzzttrra{(\applett .\the\frmcount )}\fi}%
\else\xdef\spzzttrra{(#1)}\fi%
\expandafter\xdef\csname#1\endcsname{\spzzttrra}
\eqno \hbox{\rm\spzzttrra} $$}
\def\nfr{\nameformula}    
\def\nameali#1{\global\advance\frmcount by 1%
\ifnum\draftnum=0  {\ifnum\indappcount=0%
{\ifnum\cpcount<1\xdef\spzzttrra{(\the\frmcount )}%
\else\xdef\spzzttrra{(\the\cpcount .\the\frmcount )}\fi}%
\else\xdef\spzzttrra{(\applett .\the\frmcount )}\fi}%
\else\xdef\spzzttrra{(#1)}\fi%
\expandafter\xdef\csname#1\endcsname{\spzzttrra}
  \hbox{\rm\spzzttrra} }      \clearfrmcount
\newcount\cpcount \def\clearcpcount{\global\cpcount=0}
\newcount\subcpcount \def\clearsubcpcount{\global\subcpcount=0}
\newcount\appcount \def\clearappcount{\global\appcount=0}
\newcount\indappcount \def\clearindappcount{\indappcount=0}
\newcount\sottoparcount 

\def\applett{\ifcase\appcount  \or {A}\or {B}\or {C}\or
{D}\or {E}\or {F}\or {G}\or {H}\or {I}\or {J}\or {K}\or {L}\or
{M}\or {N}\or {O}\or {P}\or {Q}\or {R}\or {S}\or {T}\or {U}\or
{V}\or {W}\or {X}\or {Y}\or {Z}\fi    \ifnum\appcount<0
\immediate\write16 {Panda ERROR - Appendix: counter "appcount"
out of range}\fi  \ifnum\appcount>26  \immediate\write16 {Panda
ERROR - Appendix: counter "appcount" out of range}\fi}
\clearappcount  \clearindappcount \newcount\connttrre
\def\clearconnttrre{\global\connttrre=0} \newcount\countref
\def\clearcountref{\global\countref=0} \clearcountref
\def\chapter#1{\global\advance\cpcount by 1 \clearfrmcount
                 \goodbreak\null\vbox{\sjump\nobreak
                 \clearsubcpcount\clearindappcount
                 \itemitem{\ttaarr\the\cpcount .\qquad}{\ttaarr #1}
                 \par\nobreak\sjump}\nobreak}
\def\section#1{\global\advance\subcpcount by 1 \goodbreak\null
               \vbox{\sjump\nobreak\ifnum\indappcount=0
                 {\ifnum\cpcount=0 {\itemitem{\ppaarr
               .\the\subcpcount\quad\enskip\ }{\ppaarr #1}\par} \else
                 {\itemitem{\ppaarr\the\cpcount .\the\subcpcount\quad
                  \enskip\ }{\ppaarr #1} \par}  \fi}
                \else{\itemitem{\ppaarr\applett .\the\subcpcount\quad
                 \enskip\ }{\ppaarr #1}\par}\fi\nobreak\jump}\nobreak}
\clearsubcpcount
\def\appendix#1{\global\advance\appcount by 1 \clearfrmcount
                  \goodbreak\null\vbox{\jump\nobreak
                  \global\advance\indappcount by 1 \clearsubcpcount
          \itemitem{ }{\hskip-40pt\ttaarr Appendix\ \applett :\ #1}
%                  \itemitem{\ttaarr App.\applett\ }{\ttaarr #1}
             \nobreak\jump\sjump}\nobreak}
\clearappcount \clearindappcount
\def\references{\goodbreak\null\vbox{\jump\nobreak
   \itemitem{}{\ttaarr References} \nobreak\jump\sjump}\nobreak}

\clearcpcount\clearcountref

\def\setchap#1{\ifnum\indappcount=0{\ifnum\subcpcount=0%
\xdef\spzzttrra{\the\cpcount}%
\else\xdef\spzzttrra{\the\cpcount .\the\subcpcount}\fi}
\else{\ifnum\subcpcount=0 \xdef\spzzttrra{\applett}%
\else\xdef\spzzttrra{\applett .\the\subcpcount}\fi}\fi
\expandafter\xdef\csname#1\endcsname{\spzzttrra}}
\newcount\draftnum \newcount\ppora   \newcount\ppminuti
\global\ppora=\time   \global\ppminuti=\time
\global\divide\ppora by 60  \draftnum=\ppora
\multiply\draftnum by 60    \global\advance\ppminuti by -\draftnum
\def\droggi{\number\day /\number\month /\number\year\ \the\ppora
:\the\ppminuti}     \global\draftnum=0
\def\draftcomment#1{\ifnum\draftnum=0 \relax \else
{\ {\bf ***}\ #1\ {\bf ***}\ }\fi} 
%
%     Maximum number of references = 200
%     boxes 50 -> 250 reserved for references
%
\catcode`@=11
\gdef\Ref#1{\expandafter\ifx\csname @rrxx@#1\endcsname\relax%
{\global\advance\countref by 1    \ifnum\countref>200
\immediate\write16 {Panda ERROR - Ref: maximum number of references
exceeded}  \expandafter\xdef\csname @rrxx@#1\endcsname{0}\else
\expandafter\xdef\csname @rrxx@#1\endcsname{\the\countref}\fi}\fi
\ifnum\draftnum=0 \csname @rrxx@#1\endcsname \else#1\fi}
\gdef\beginref{\ifnum\draftnum=0  \gdef\Rref{\fairef}
\gdef\endref{\scriviref} \else\relax\fi
\ifx\risposta\mplarisposta \ninepoint \fi
\parskip 2pt plus.2pt \baselineskip=12pt}
\def\Reflab#1{[#1]} \gdef\Rref#1#2{\item{\Reflab{#1}}{#2}}
\gdef\endref{\relax}  \newcount\conttemp
\gdef\fairef#1#2{\expandafter\ifx\csname @rrxx@#1\endcsname\relax
{\global\conttemp=0 \immediate\write16 {Panda ERROR - Ref: reference
[#1] undefined}} \else
{\global\conttemp=\csname @rrxx@#1\endcsname } \fi
\global\advance\conttemp by 50  \global\setbox\conttemp=\hbox{#2} }
\gdef\scriviref{\clearconnttrre\conttemp=50
\loop\ifnum\connttrre<\countref \advance\conttemp by 1
\advance\connttrre by 1
\item{\Reflab{\the\connttrre}}{\unhcopy\conttemp} \repeat}
\clearcountref \clearconnttrre
\catcode`@=12
\ifx\risposta\mplarisposta \def\Reflab#1{#1.} \letterfootnote \fi

\def\slashchar#1{\setbox0=\hbox{$#1$} \dimen0=\wd0
     \setbox1=\hbox{/} \dimen1=\wd1 \ifdim\dimen0>\dimen1
      \rlap{\hbox to \dimen0{\hfil/\hfil}} #1 \else
      \rlap{\hbox to \dimen1{\hfil$#1$\hfil}} / \fi}
\ifx\oldchi\undefined \let\oldchi=\chi
  \def\cchi{{\raise 1pt\hbox{$\oldchi$}}} \let\chi=\cchi \fi

\def\frac#1#2{{\textstyle{#1 \over #2}}}

\def\half{\ifinner {\scriptstyle {1 \over 2}}\else {1 \over 2} \fi}

\def\simge{\rlap{\raise 2pt \hbox{$>$}}{\lower 2pt \hbox{$\sim$}}}
\def\simle{\rlap{\raise 2pt \hbox{$<$}}{\lower 2pt \hbox{$\sim$}}}

\def\vbig#1#2{{\vbigd@men=#2\divide\vbigd@men by 2%
\hbox{$\left#1\vbox to \vbigd@men{}\right.\n@space$}}}

%
%--------------------------------------------------------------------
%
\newcount\fdefcontre \newcount\fdefcount \newcount\indcount
\newread\filefdef  \newread\fileftmp  \newwrite\filefdef
\newwrite\fileftmp     \def\strip#1*.A {#1}
\def\futuredef#1{\beginfdef
\expandafter\ifx\csname#1\endcsname\relax%
{\immediate\write\fileftmp {#1*.A}
\immediate\write16 {Panda Warning - fdef: macro "#1" on page
\the\pageno \space undefined}
\ifnum\draftnum=0 \expandafter\xdef\csname#1\endcsname{(?)}
\else \expandafter\xdef\csname#1\endcsname{(#1)} \fi
\global\advance\fdefcount by 1}\fi   \csname#1\endcsname}

\def\beginfdef{\ifnum\fdefcontre=0
\immediate\openin\filefdef \jobname.fdef
\immediate\openout\fileftmp \jobname.ftmp
\global\fdefcontre=1  \ifeof\filefdef \immediate\write16 {Panda
WARNING - fdef: file \jobname.fdef not found, run TeX again}
\else \immediate\read\filefdef to\spzzttrra
\global\advance\fdefcount by \spzzttrra
\indcount=0      \loop\ifnum\indcount<\fdefcount
\advance\indcount by 1   \immediate\read\filefdef to\spezttrra
\immediate\read\filefdef to\sppzttrra
\edef\spzzttrra{\expandafter\strip\spezttrra}
\immediate\write\fileftmp {\spzzttrra *.A}
\expandafter\xdef\csname\spzzttrra\endcsname{\sppzttrra}
\repeat \fi \immediate\closein\filefdef \fi}
\def\endfdef{\immediate\closeout\fileftmp   \ifnum\fdefcount>0
\immediate\openin\fileftmp \jobname.ftmp
\immediate\openout\filefdef \jobname.fdef
\immediate\write\filefdef {\the\fdefcount}   \indcount=0
\loop\ifnum\indcount<\fdefcount    \advance\indcount by 1
\immediate\read\fileftmp to\spezttrra
\edef\spzzttrra{\expandafter\strip\spezttrra}
\immediate\write\filefdef{\spzzttrra *.A}
\edef\spezttrra{\string{\csname\spzzttrra\endcsname\string}}
\iwritel\filefdef{\spezttrra}
\repeat  \immediate\closein\fileftmp \immediate\closeout\filefdef
\immediate\write16 {Panda Warning - fdef: Label(s) may have changed,
re-run TeX to get them right}\fi}
\def\iwritel#1#2{\newlinechar=-1
{\newlinechar=`\ \immediate\write#1{#2}}\newlinechar=-1}
\global\fdefcontre=0 \global\fdefcount=0 \global\indcount=0
%
%--------------------------------------------------------------------
%
\null
%
%--------------------------------------------------------------------
%
%                             THE    END
%
%--------------------------------------------------------------------
%